# Monotonic decrease of the quantum nonadditive divergence by projective measurements

Sumiyoshi Abe

*Institute of Physics, University of Tsukuba,*

*Ibaraki 305-8571, Japan*

Nonadditive (nonextensive) generalization of the quantum Kullback-Leibler divergence, termed the quantum $q$-divergence, is shown not to increase by projective measurements in an elementary manner.

In recent papers [1-3], we have developed a nonadditive generalization of information theory and have discussed its distinguished roles in the study of quantum entanglement extensively (see also, [4-8]). These works have primarily been concerned with the Tsallis nonadditive (nonextensive) entropy [9] and the associated generalized conditional entropy [1]. On the other hand, quite recently, the role of the generalized Kullback-Leibler divergence, termed the quantum $q$-divergence, has been examined as a measure of the degree of state purification [10]. There, an advantageous point of the quantum $q$-divergence over the ordinary quantum Kullback-Leibler divergence has been clarified [see the discussion after Eq. (5) below].

In this article, we study the behavior of the quantum $q$-divergence under measurements, i.e., quantum operations. In particular, we present an elementary proof that the quantum $q$-divergence does not increase by projective measurements.

The quantum $q$-divergence is the relative entropy associated with the Tsallis entropy. The Tsallis entropy reads

$$S_q[\rho] = -\text{Tr}\left(\rho^q \ln_q \rho\right). \tag{1}$$

Here, $\rho$ is the normalized density matrix of the quantum system under consideration and $q$ is the entropic index which can be an arbitrary positive number at this level. $\ln_q x$ stands for the $q$-logarithmic function [11] defined by $\ln_q x = (x^{1-q} - 1)/(1-q)$, which tends to the ordinary logarithmic function, $\ln x$, in the limit $q \to 1$. Then, the quantum $q$-divergence of $\rho$ with respect to the reference density matrix $\sigma$ is given by

$$K_q[\rho \| \sigma] = \mathrm{Tr}\left[\rho^q \left(\ln_q \rho - \ln_q \sigma\right)\right]. \tag{2}$$

(The classical counterpart of this quantity has been introduced independently and almost simultaneously in [12-14].) Using the definition of the $q$-logarithmic function, Eq. (2) can also be written in the following compact form:

$$K_q[\rho \| \sigma] = \frac{1}{1-q}\left[1 - \mathrm{Tr}\left(\rho^q \sigma^{1-q}\right)\right]. \tag{3}$$

Since this quantity should not be too sensitive to small eigenvalues of the density matrices, the range of $q$ is taken to be

$$0 < q < 1. \tag{4}$$

Let $s_\rho$ and $s_\sigma$ be the supports of $\rho$ and $\sigma$, respectively. In the case when $s_\rho \leq s_\sigma$, $K_q[\rho \| \sigma]$ has the well-defined limit $q \to 1-0$, which yields the ordinary quantum Kullback-Leibler divergence introduced by Umegaki [15]

$$K[\rho \| \sigma] = \text{Tr}\big[\rho(\ln\rho - \ln\sigma)\big]. \tag{5}$$

Here, the condition, $s_\rho \leq s_\sigma$, is crucial. In fact, $K[\rho \| \sigma]$ becomes singular when $s_\rho > s_\sigma$. Therefore, $K[\rho \| \sigma]$ cannot be defined if, for example, $\sigma$ is a pure state (i.e., an idempotent operator), since $\ln\sigma = (\sigma - I)\zeta(1)$, which is divergent, where $I$ and $\zeta(s)$ are the identity operator and the Riemann zeta function, respectively. In marked contrast to this, $K_q[\rho \| \sigma]$ with $q \in (0, 1)$ remains well-defined even in such a case [10].

In Ref. [10], it has been shown that (i) $K_q[\rho \| \sigma] \geq 0$ and $K_q[\rho \| \sigma] = 0$ if and only if $\rho = \sigma$, (ii) for product states, $\rho(A, B) = \rho_1(A) \otimes \rho_2(B)$ and $\sigma(A, B) = \sigma_1(A) \otimes \sigma_2(B)$, of a bipartite system $(A, B)$, $K_q[\rho \| \sigma]$ satisfies pseudoadditivity: $K_q[\rho_1 \otimes \rho_2 \| \sigma_1 \otimes \sigma_2] = K_q[\rho_1 \| \sigma_1] + K_q[\rho_2 \| \sigma_2] + (q-1)K_q[\rho_1 \| \sigma_1]K_q[\rho_2 \| \sigma_2]$ and (iii) $K_q$ can be observed as the $q$-analog (i.e., $q$-deformation) of $K$ in the sense in [16].

In addition to the properties (i)-(iii), we wish to notice another important one anew here. That is, $K_q[\rho \| \sigma]$ is jointly convex

$$K_q\Big[\sum_i \lambda_i \rho^{(i)} \| \sum_i \lambda_i \sigma^{(i)}\Big] \leq \sum_i \lambda_i K_q\big[\rho^{(i)} \| \sigma^{(i)}\big], \tag{6}$$

where $\lambda_i > 0$ and $\sum_i \lambda_i = 1$. This directly follows from the expression in Eq. (3) as well as Lieb's theorem [17] stating that $\text{Tr}\big(L^{1-x} M^x\big)$ with $x \in (0, 1)$ is jointly concave in any positive operators, *L* and *M*. Eq. (6) generalizes joint convexity of the ordinary quantum divergence (see [18], for example).

Now, let us discuss the behavior of $K_q[\rho \| \sigma]$ under projective measurement of $\rho$ and $\sigma$. This measurement can be regarded as a particular kind of positive trace-preserving quantum operation, but is quite common from the experimental viewpoint [19]. Let $Q$ be an observable with eigenspaces defined by orthogonal projections $P_k$ and $\{q_k\}$ be its measured values. Then, $Q = \sum_k q_k P_k$, $P_k P_{k'} = \delta_{kk'} P_{k'}$ and $\sum_k P_k = I$. The finite probability $p_k$ of obtaining the value $q_k$ of $Q$ in a state $\rho$ of the system through the projective measurement is $p_k = \text{Tr}(\rho P_k)$. From this, $\rho$ is transformed to $\rho_k = p_k^{-1} P_k \rho P_k$. Averaging over all possible outcomes, we have

$$\Pi(\rho) = \sum_k p_k \rho_k = \sum_k P_k \rho P_k. \qquad (7)$$

Clearly, $\Pi$ is a positive trace-preserving operation.

Let us employ the diagonal representations of $\rho$ and $\sigma$:

$$\rho = \sum_a r(a) |a\rangle\langle a|, \qquad \sigma = \sum_b s(b) |b\rangle\langle b|, \qquad (8)$$

where $r(a) \geq 0$, $\sum_a r(a) = 1$, $\langle a|a'\rangle = \delta_{aa'}$, $\sum_a |a\rangle\langle a| = I$ and so on. Under the operation of a projective measurement, they are replaced by

$$\Pi(\rho) = \sum_{a} r(a)\,\Pi(|a\rangle\langle a|), \qquad \Pi(\sigma) = \sum_{b} s(b)\,\Pi(|b\rangle\langle b|), \qquad (9)$$

respectively. Let us further use the diagonal representations

$$\Pi(|a\rangle\langle a|) = \sum_{\alpha} \mu(\alpha, a)\,|\alpha\rangle\langle\alpha|, \qquad \Pi(|b\rangle\langle b|) = \sum_{\beta} \nu(\beta, b)\,|\beta\rangle\langle\beta|, \qquad (10)$$

where $\mu(\alpha, a) = \sum_{k} |\langle\alpha|P_k|a\rangle|^2 \geq 0$, $\sum_{a} \mu(\alpha, a) = \sum_{\alpha} \mu(\alpha, a) = 1$, $\langle\alpha|\alpha'\rangle = \delta_{\alpha\alpha'}$, $\sum_{\alpha} |\alpha\rangle\langle\alpha| = I$ and so on. These decompositions in Eq. (10) are valid if the projection operator is identified with $P_{k=\alpha} = |\alpha\rangle\langle\alpha|$. Henceforth, the projection operator is simply written as $P_k = |k\rangle\langle k|$, and accordingly Eq. (10) may be reexpressed as follows:

$$\Pi(|a\rangle\langle a|) = \sum_{k} \mu(k, a)\,|k\rangle\langle k|, \quad \Pi(|b\rangle\langle b|) = \sum_{k} \nu(k, b)\,|k\rangle\langle k|, \qquad (11)$$

where $\mu(k, a) = |\langle k|a\rangle|^2$, $\nu(k, b) = |\langle k|b\rangle|^2$, $\langle k|k'\rangle = \delta_{kk'}$ and $\sum_{k} |k\rangle\langle k| = \sum_{k} P_k = I$. Therefore, we have

$$[\Pi(\rho)]^q = \sum_{k} \left[\sum_{a} r(a)\,\mu(k, a)\right]^q P_k,$$

$$[\Pi(\sigma)]^{1-q} = \sum_k \left[ \sum_b s(b)\, \nu(k,b) \right]^{1-q} P_k, \qquad (12)$$

which lead to

$$\mathrm{Tr}\left\{ [\Pi(\rho)]^q\, [\Pi(\sigma)]^{1-q} \right\} = \sum_k \left[ \sum_a r(a)\, \mu(k,a) \right]^q \left[ \sum_b s(b)\, \nu(k,b) \right]^{1-q}. \qquad (13)$$

Since $f(x) = x^p$ ($x > 0$, $0 < p < 1$) is a concave function, holds $f\left( \sum_i \lambda_i a_i \right) \geq \sum_i \lambda_i f(a_i)$ for $\lambda_i > 0$ and $\sum_i \lambda_i = 1$. Therefore, it follows that

$$\mathrm{Tr}\left\{ [\Pi(\rho)]^q\, [\Pi(\sigma)]^{1-q} \right\}$$

$$\geq \sum_k \left( \sum_a \mu(k,a)\, [r(a)]^q \right) \left( \sum_b \nu(k,b)\, [s(b)]^{1-q} \right)$$

$$= \sum_{a,b} [r(a)]^q [s(b)]^{1-q}\, \mathrm{Tr}\left[ \Pi(|a\rangle\langle a|)\, |b\rangle\langle b| \right]. \qquad (14)$$

So far, no peculiar assumptions have been made on the algebraic structure of $\Pi$ in connection with $\sigma$. To the best of our knowledge, to proceed further, it seems necessary to assume that $\Pi$ is an "expectation" [18]: $\Pi(|a\rangle\langle a|)\, |b\rangle\langle b| = \Pi(|a\rangle\langle a|b\rangle\langle b|)$. This essentially implies that $P_k$ ($\forall\, k$) can commute with $\sigma$. Then, Eq. (14) yields

$$\mathrm{Tr}\left\{[\Pi(\rho)]^q \, [\Pi(\sigma)]^{1-q}\right\} \geq \mathrm{Tr}\left(\rho^q \, \sigma^{1-q}\right), \tag{15}$$

leading to

$$K_q[\Pi(\rho)\|\Pi(\sigma)] \leq K_q[\rho\|\sigma]. \tag{16}$$

Consequently, we obtain the main result that the quantum $q$-divergence does not increase by projective measurements.

In conclusion, we have shown that the quantum $q$-divergence is jointly convex and does not increase by projective measurements. Physically, this implies (Tsallis) entropy production by the measurements. Quite recently, it has been shown [20] that Clausius' inequality can be established in nonextensive quantum thermodynamics by making use of the quantum $q$-divergence and its monotonicity with respect to trace-preserving completely positive unital quantum operations. Further investigation in this direction may be important for developing thermodynamics of small systems [20].